\pdfoutput=1
\documentclass[preprint,12pt, 5p, twocolumn]{elsarticle}
\usepackage{amssymb}
\usepackage{amsmath}
\usepackage{threeparttable}
\usepackage{makecell}
\usepackage{graphicx}
\usepackage{multirow}
\usepackage{booktabs}
\usepackage{bm}
\usepackage{array}
\usepackage{float}
\usepackage{booktabs}

\newtheorem{definition}{Definition}
\newtheorem{example}{Example}

\begin{document}

\begin{frontmatter}

\title{Dynamic Network Embedding Survey}


\author[1]{Guotong Xue}
\author[1]{Ming Zhong}
\author[2]{Jianxin Li}
\author[1]{Jia Chen}
\author[1]{Chengshuai Zhai}
\author[3]{Ruochen Kong}
\address[1]{School of Computer Science, Wuhan University, Wuhan, China}
\address[2]{School of Information Technology, Deakin University, Melbourne, Australia}
\address[3]{School of Computer Science, Rensselaer Polytechnic Institute, New York, USA}

\begin{abstract}
    Since many real world networks are evolving over time, such as social networks and user-item networks, there are increasing research efforts on dynamic network embedding in recent years.
    They learn node representations from a sequence of evolving graphs but not only the latest network, for preserving both structural and temporal information from the dynamic networks.
    Due to the lack of comprehensive investigation of them, we give a survey of dynamic network embedding in this paper.
    Our survey inspects the data model, representation learning technique, evaluation and application of current related works and derives common patterns from them.
    Specifically, we present two basic data models, namely, discrete model and continuous model for dynamic networks.
    Correspondingly, we summarize two major categories of dynamic network embedding techniques, namely, structural-first and temporal-first that are adopted by most related works.
    Then we build a taxonomy that refines the category hierarchy by typical learning models. The popular experimental data sets and applications are also summarized. Lastly, we have a discussion of several distinct research topics in dynamic network embedding.
\end{abstract}


\end{frontmatter}


\section{Introduction}
In the last few years, the static network embedding problem has been intensively studied, and a variety of techniques have been proposed to learn network topology and attributes\cite{roweis2000nonlinear,wang2016structural,tang2015line,perozzi2014deepwalk,defferrard2016convolutional,DBLP:journals/isci/ChenQ20,tcyb2021/mzhong}. 
However, the networks are dynamic, namely, evolving over time in the real world. 
Not only the node and edge attributes may change as time goes by, but more importantly, the topology would also change, like creating or removing nodes and edges. Let us consider the following real world dynamic networks as examples.

\begin{figure}[!t]
    \centering
    \includegraphics[width = .5\textwidth]{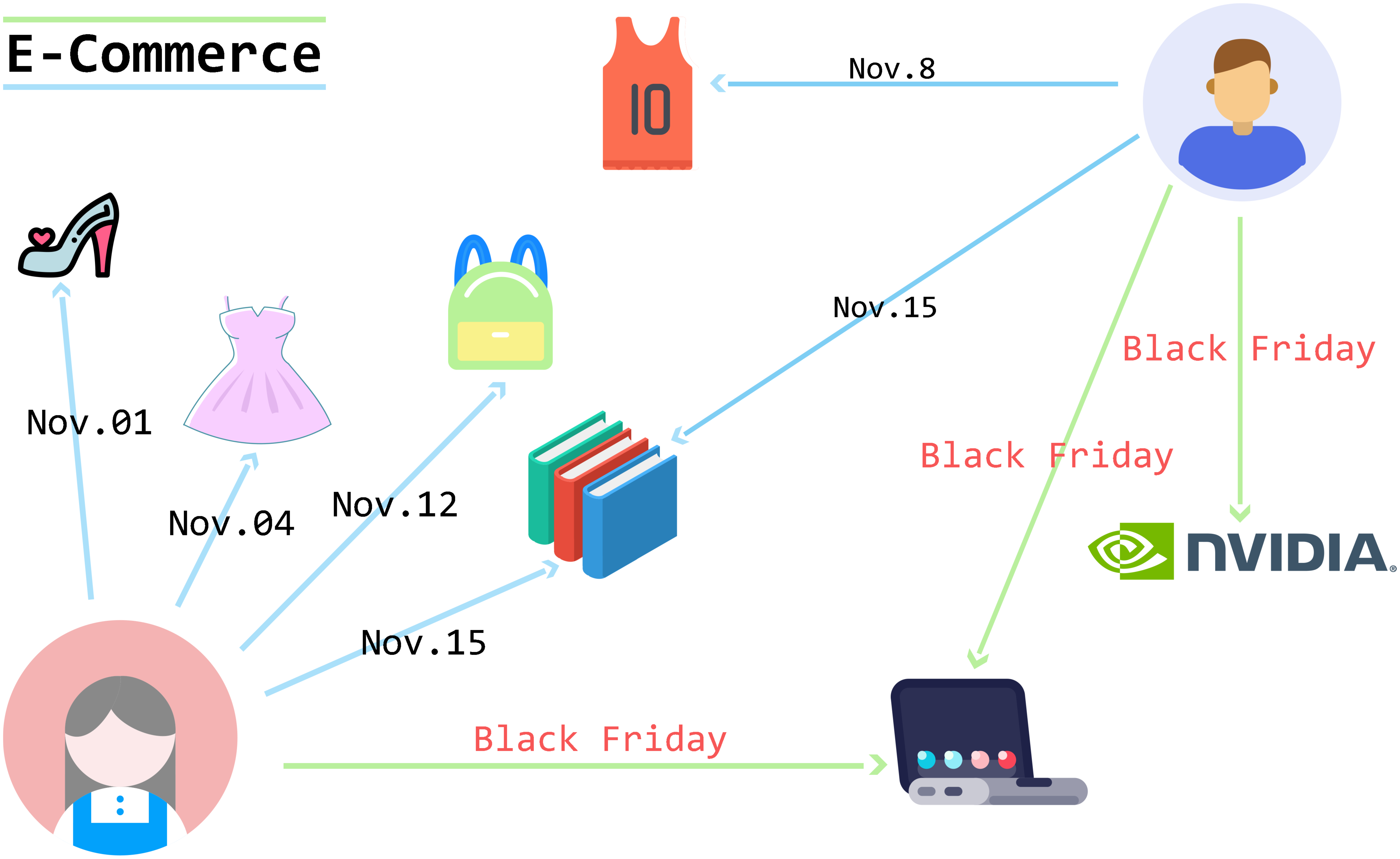}
    \caption{An illustration of user-item graph.}
    \label{pic:ecommerce}
\end{figure}

\begin{example}[User-item network]
    The user-item network illustrated in Figure~\ref{pic:ecommerce} is usually used for recommendation in e-commerce.
    It is recently noticed that a user would interact with different kinds of item in different periods, which formulates different temporary patterns in the network.
    For instance, a bunch of bursty edges may be created due to the sales on special shopping events.
    Therefore, the recommendation engine should recognize such pattern in a particular period.
\end{example}

\begin{example}[Payment network]
    The payment network is built from the electronic payment transactions in banks, credit card companies, and 3rd party payment institutions.
    Obviously, the payment network has constantly incoming nodes and edges every second.
    It has been leveraged to identify abnormal activities like financial fraud and money laundry in a real time fashion.
\end{example}

\begin{example}[Social Network]
    The social network is another highly dynamic network that evolves with online activities on social platforms.
    Normally, a user may have changing social interests, and thus may fall into different so-called short-term communities in the network.
    In order to better understand the evolving nature of human interactions, the social network analytics like link prediction and community detection should take the temporal information into consideration.
\end{example}

Motivated by the essential dynamics of real-world networks, learning node representation for dynamic networks has become a more and more popular research topic recently.
Fundamentally, the learned dynamic network embeddings should capture the network structure and reflect the temporal evolution. 
Namely, the learned embeddings should not only maintain the structural relationships between nodes in vector space, but also is required to describe the topological changes.
However, it's hard to learn embeddings on dynamic network. We summarize the challenges as follows: 
Firstly, since the edges and nodes are evolving over time, temporal constraints are imposed on neighborhood aggregation methods. 
Besides, in order to model the dynamic of network, the learned node embeddings should also be functions of time to encode the temporal information. For example, considering that the time of node interactions are different, the impacts are also different.
Furthermore, efficient learning methods are important since the scale of network may grow over time.

Though facing challenges above, the dynamic network embedding has the following significance compared with the static network embedding.

\begin{itemize}
    \item
    \textbf{Capture temporal information}
    The temporal information is helpful for accurately analyzing network properties~\cite{newman2001structure, watts1998collective}.
    However, static network embedding methods are not capable of dealing with it.
    Besides, embedding vectors may be reformed into different spaces if we learn them by rerunning static model several times.
    With learned dynamic network embeddings, researchers can utilize the additional information and better understand the network evolving process.

    \item
    \textbf{Update representation in a fine-grained granularity of time.}
    Instead of accumulating the historical changes of network and integrating them into a single static network, dynamic network embedding methods represent dynamic network as a series of snapshots or a sequence of new node/edge with timestamp.
    In this way, the node representation can be updated at the required granularity of time.
    What's more, with the support of RNN series models, the embeddings could be updated whenever a new interaction occurs.
    This could be applied in the scenarios requiring real-time performance.


    \item
    \textbf{Achieve high efficiency of updating representation.}
    The straightforward way of refreshing representation for dynamic networks is simply applying a static network embedding approach from scratch at each time step.
    However, retraining a network embedding model could be very time-consuming and cost high resources, especially for large scale networks.
    In contrast, most dynamic network embedding models could avoid retraining from scratch and achieve considerable update efficiency.

\end{itemize}

In this paper, we survey the state-of-the-art dynamic network embedding approaches.
To the best knowledge we have, there is one published survey\cite{kazemi2020representation} and a preprint on arxiv.org\cite{xie2020survey} in this field.
In contrast to them, our survey aims to
1) inspect the related works from more perspectives like data models,
2) build the taxonomy of existing techniques from higher abstract level based on underlying data models,
and 3) have more discussion of common and principal issues involved in the related work like out-of-sample node embedding and prediction of future embedding.
Compared with \cite{kazemi2020representation}, our survey hopes to serve the researchers who have good knowledges of network embedding and just wish to know the recent progress on dynamic networks with more concise description and more practical taxonomy.
Therefore, besides the summarization of related work, our survey would organize and compare them in several important aspects,
and derive general and significant patterns from them, in order to assist the future work on dynamic network embedding.

Our contributions are summarized as follows.

\begin{itemize}
    \item
    We present two common network data models underlying existing dynamic network embedding approaches, and propose a taxonomy that organizes these approaches by high-level methodologies and then by concrete learning models. The two methodologies summarized by us are actually two natural ways of handling the two data models respectively.
    \item
    We conduct comprehensive review of recent progress in dynamic network embedding, and make necessary comparison between them.
    \item
    We discuss several distinct and critical challenges in dynamic network embedding and explain how relevant existing approaches address them.
    \item
    We summarize the useful resources in existing approaches, like experimental datasets, evaluation tasks and real applications of dynamic network embedding.
\end{itemize}

The rest of this article is organized as follows.
Section~\ref{sec:model} presents the common data models used by related works.
Section~\ref{sec:taxonomy} gives a taxonomy of embedding techniques used by related work.
Section~\ref{sec:discussion} conducts discussion of challenging issues.
Section~\ref{sec:application} introduces commonly used datasets, evaluation, and practical applications.

\section{Dynamic Network Models}
\label{sec:model}

In this section, we will introduce the data models of dynamic networks.
Unlike the static network embedding approaches that almost follow a uniform network data model, the dynamic network embedding approaches have quite different definitions of dynamic network, which have significant impact on the selection of embedding techniques (as discussed in Section~\ref{sec:taxonomy}).
Therefore, it is necessary to investigate the dynamic network models.

Basically, a static network is defined as a (attributed) network.

\begin{definition}[Network]
    A (attributed) network is represented as $G = (V;E;A)$, where $V$ is a set of vertices or nodes, $E \subseteq V \times V$ is a set of edges between the nodes,
    and optionally, for each node $v \in V$, $A(v) \in \mathbb{R}^n$ is an attribute vector.
\end{definition}

The dynamic networks are graphs that have nodes, edges and attributes updated gradually over time.
Naturally, there are two ways to update graphs, namely, discrete updates time interval by time interval and continuous updates in a stream.
Correspondingly, there are two typical variants of the network data model that represent the dynamic networks.
Next, we formally present these two models respectively.

\subsection{Discrete Model}

\begin{figure}[ht]
    \centering
    \includegraphics[width = .4\textwidth]{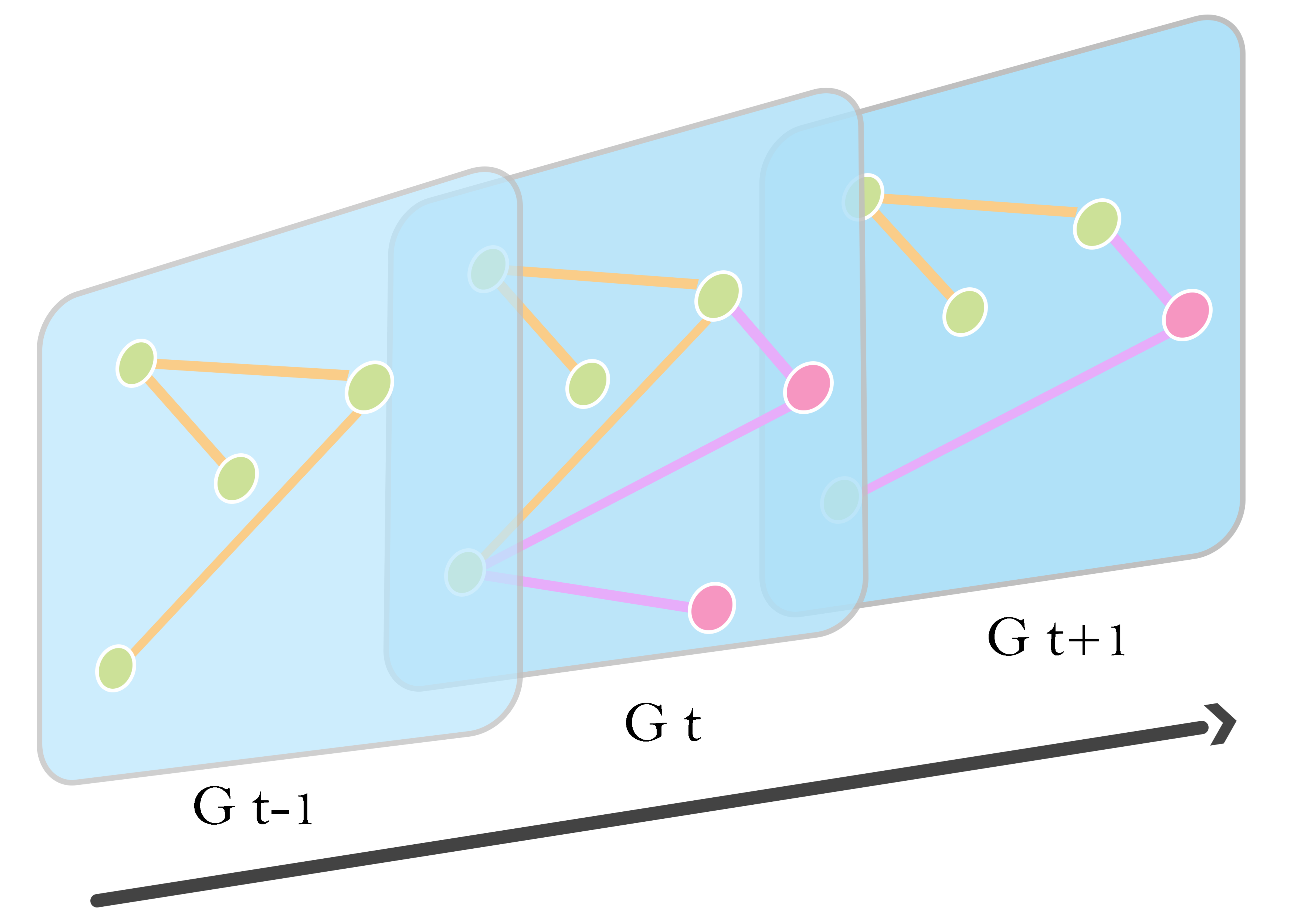}
    \caption{An illustration of discrete model.}
    \label{pic:discrete}
\end{figure}

Most related works\cite{DANE,Timers,DHPE,stgcn,extendedskipgram,DynGEM,DepthLGP,DGNN,dynnode2vec,dysat,DynamicTriad,Evolvegcn,BurstGraph,dyngraph2vec,dyhan}
use the discrete model to represent the dynamic networks.
Intuitively, they see a dynamic network as a sequence of network snapshots, namely, independent and complete graphs.
Normally, as figure~\ref{pic:discrete} showed, the structure of graphs changes a bit between adjacent snapshots, which means a few of nodes and edges will be created or removed between snapshots.

We formally define the discrete model as follows.

\begin{definition}[Discrete Model]
    A discrete model of dynamic network $G$ is a sequence of network snapshots within a given time interval.
    We have $G=\{G_1,\dots,G_T\}$, where $T$ is the number of snapshots.
    Each snapshot $G_t=(V_t, E_t)$ is a static network recorded at time $t$.
\end{definition}

\subsection{Continuous Model}

\begin{figure}[ht]
    \centering
    \includegraphics[width = .45\textwidth]{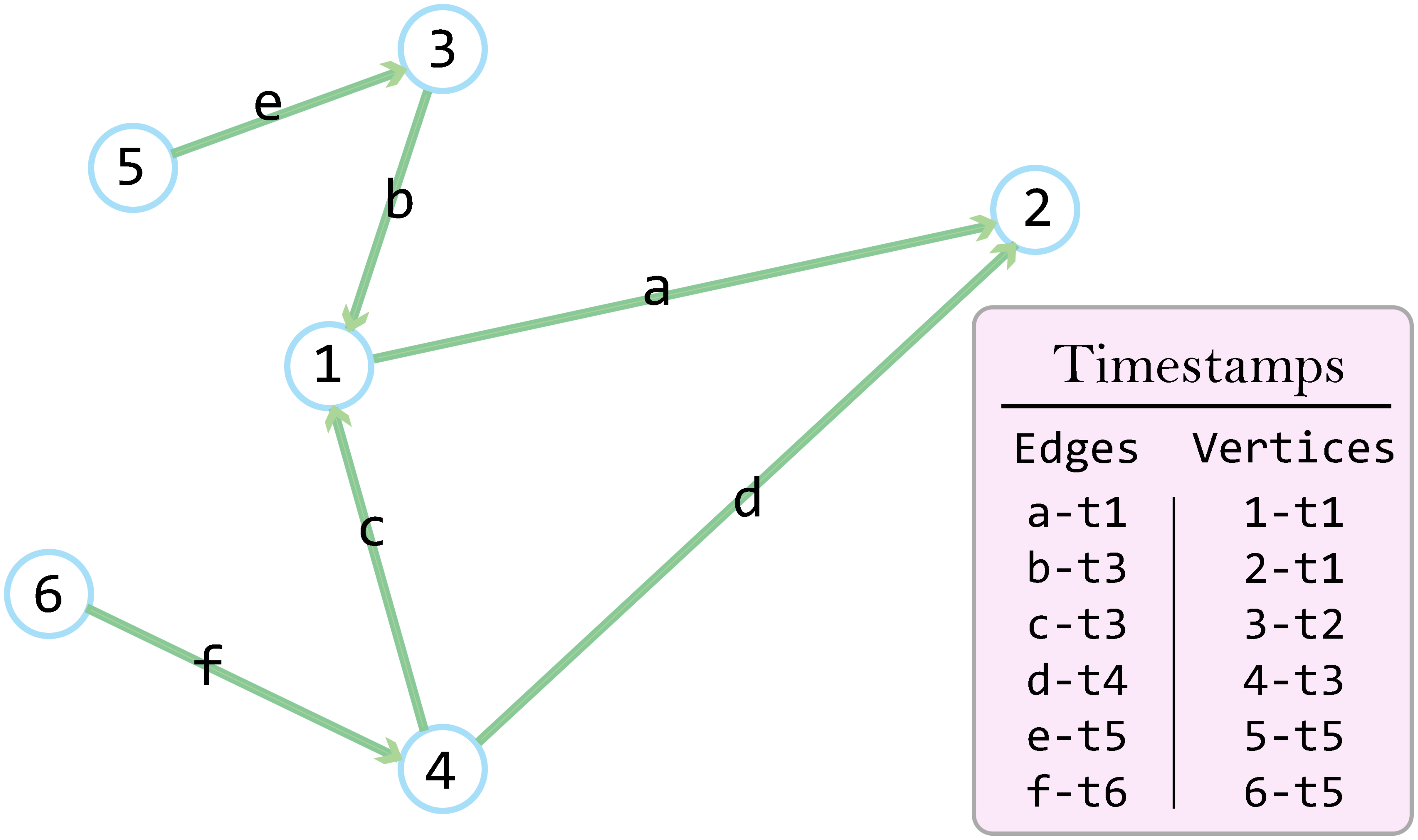}
    \caption{An illustration of continuous model.}
    \label{pic:continuous}
\end{figure}

In contrast, other related works\cite{ContinuoustimeDNE,knowevolve,Dyrep,Netwalk,htne,jodie,tgat} use the continuous model, in which new nodes and edges are added in a stream manner.
New edges are created and annotated with timestamps when interactions between nodes occur, which is illustrated in figure~\ref{pic:continuous}.
Also, all nodes update their own timestamps when they are created or their properties changed.

We formally defined the continuous model as follows.

\begin{definition}[Continuous Model]
    A continuous model of dynamic network $G$ is a network with edges and ndoes annotated with timestamps.
    We have $G=(V_T,E_T, \mathcal{T})$ where $\mathcal{T}: V,E \rightarrow \mathbb{R}^{+}$ is a function that maps each edge and node to a corresponding timestamp.
\end{definition}

In addition, it should be noted that the exactly used network model could be slightly different between each approach.
For example, some methods consider network with attributes\cite{stgcn,knowevolve,DepthLGP,DANE,dysat,Dyrep,htne,jodie,Evolvegcn,AddGraph,BurstGraph,tgat,dyhan} while others\cite{DGNN,extendedskipgram,ContinuoustimeDNE,Netwalk,DynamicTriad,dynnode2vec,dyngraph2vec} only deal with a raw network.
What's more, as we summarized in table~\ref{table:summary}, there are some methods that can only tackle with adding nodes in the evolution while leave the deletion of nodes in future development.

\section{Taxonomy of Embedding Techniques}
\label{sec:taxonomy}

In this section, we present a taxonomy of dynamic network embedding techniques. To explain the design of our taxonomy, it is needed to clarify the following fundamental challenges of learning dynamic network embedding firstly.

\begin{itemize}
    \item \textbf{Preserve the networks' structure (and properties).}
    For the structure of the network, the learned embeddings should reflect the neighbor relationship between nodes and preserve high-order proximity. For example, nodes in same community should have similar embeddings . 
    \item \textbf{Preserve the evolution of networks in long-term.}
    For the evolution of network, the learned embeddings should reflect the interaction with other nodes over time. 
\end{itemize}

To address both the above challenges in one shot, there are naturally two highly abstract methodologies.
Firstly, import the temporal features of dynamic network into the learning models that originally preserve the structure (and property) information of static networks, such as skip-gram\cite{ContinuoustimeDNE,htne,dynnode2vec,extendedskipgram}, auto-encoder\cite{DynGEM,Netwalk,BurstGraph,dyngraph2vec}, GNN\cite{stgcn,dysat,dyhan,Evolvegcn,AddGraph}, etc.
We call this methodology as structural-first.
Secondly, import the structure (and property) features of static networks into the learning model that originally preserve the temporal information of dynamical data, such as RNN, LSTM\cite{jodie,knowevolve,Dyrep,DGNN}, etc.
We call this methodology as temporal-first.

In fact, most related works adopt these two methodologies. Therefore, we categorize the dynamic network embedding techniques into three categories: structural-first, temporal-first, and others.
In each category, there would be more specific sub-categories corresponding to different learning models.
For example, Skip-gram based model, matrix factorization based model, auto-encoder based model and GNN based model are the sub-categories of structural-first.
The complete taxonomy is shown in table~\ref{table:summary}.

\begin{table*}[ht]
    \begin{threeparttable}
        \centering
        \resizebox*{\textwidth}{100mm}{
            \begin{tabular}{*{12}{c}}
                \toprule
                \multirow{2}*{Method} & \multirow{2}*{Methodology} & \multicolumn{3}{c}{Data Mode} & \multicolumn{4}{c}{Learning Techniques} & \multicolumn{3}{c}{Experimental tasks} \\
                \cmidrule(lr){3-5} \cmidrule(lr){6-9} \cmidrule(lr){10-12}
                & & Model Type & \makecell[c]{Node/Edge \\ Addition}  & \makecell[c]{Node/Edge \\Deletion} & \makecell[c]{Matrix\\Factorization} & Skip-gram & \makecell[c]{Deep\\Learning} & Others & \makecell[c]{Node \\ Classification} & \makecell[c]{Link \\ Prediction} & Others \\

                \midrule
                DANE\cite{DANE} & S-first & discrete & both & both & yes & & & & yes & & network clustering \\
                DHPE\cite{DHPE} & S-first & discrete & both & both & yes & & & & yes & yes & \makecell[c]{high-order proximity \\approximation} \\
                Know-Evolve\cite{knowevolve} & T-first & continuous & both & & & & yes & & & yes & time prediction \\
                STGCN\cite{stgcn} & S-first & discrete & & & & & yes & & & & traffic prediction \\
                DNE\cite{extendedskipgram} & S-first & discrete & both & both & & yes & & & yes & & network layout \\
                DynGEM\cite{DynGEM} & S-first & discrete & both & both & & & yes & & & yes & network reconstruction \\
                DepthLGP\cite{DepthLGP} & Others & discrete & both & & & & & yes & yes & yes & \\
                DGNN\cite{DGNN} & T-first & discrete & both & & & & yes & & yes & yes & \\
                dynnode2vec\cite{dynnode2vec} & S-first & discrete & both & both & & yes & & & yes & yes & anomaly detection \\
                \makecell[c]{Continuous-Time \\Dynamic Networks\cite{ContinuoustimeDNE}} & S-first & continuous & edge & & & yes  & & & & yes & \\
                DySAT\cite{dysat} & S-first & discrete & edge & edge & & &yes & & & yes & \\
                DyRep\cite{Dyrep} & T-first & continuous & both & & & & yes & & & yes & time prediction \\
                Netwalk\cite{Netwalk} & S-first & continuous & edge & edge & & & yes & & & & anomaly detection \\
                DynamicTriad\cite{DynamicTriad} & Others & discrete & both & & & & & yes & yes & yes & link reconstruction \\
                HTNE\cite{htne} & S-first & continuous & edge & & & yes & & & yes & yes & \\
                JODIE\cite{jodie} & T-first & continuous & edge & & & & yes & & & yes & \makecell[c]{user state\\ change prediction} \\
                EvolveGCN\cite{Evolvegcn} & S-first & discrete & both & & & & yes & & yes & yes & edge classification \\
                BurstGraph\cite{BurstGraph} & S-first & discrete & both & & & & yes& & & yes & \\
                AddGraph\cite{AddGraph} & S-first & discrete & edge & & & & yes & & & & anomaly detection \\
                dyngraph2vec\cite{dyngraph2vec} & S-first & discrete & edge & edge & & & yes & & & yes & \\
                TGAT\cite{tgat} & Others & discrete & both & & & & yes & & yes & yes & \\
                DyHAN\cite{dyhan} & S-first & discrete & both & both & & & yes & & & yes & \\
                DyHATR\cite{DyHATR} & T-first & discrete & both & both & & & yes & & & yes & \\
                THIGE\cite{THIGE} & T-first & continuous & edge & edge & & & yes & & & & next-item recommendation \\
                MMDNE\cite{MMDNE} & Others & continuous & both & & & & yes & yes & yes & yes & network reconstruction\\
                \bottomrule

            \end{tabular}}
            \begin{tablenotes}
                \item \textbf{T-first} means temporal first while \textbf{S-first} means structural first.
            \end{tablenotes}
        \caption{A Summary of Dynamic Network Embedding Methods}
        \label{table:summary}
    \end{threeparttable}
\end{table*}

\subsection{Structural-first model}

\subsubsection{Matrix factorization based model}

Matrix factorization techniques are one of the most traditional but effective ways to learn network embedding\cite{cai2010graph, cao2015grarep}.
We usually hope to find some lower dimensional matrices whose product serves as a approximation to the original matrix.

Matrix factorization based models often consider dynamic network as a constant changing matrices. Therefore, they could derive the embedding vectors through generalized SVD.
Moreover, the matrix perturbation theory\cite{stewart1990matrix} is also adopted to update the embedding results.

\cite{DANE} considers the raw representations of network topology and node attributes are different. Either of these could be incomplete and noisy.
Therefore, they propose a dynamic attributed network embedding framework \textit{DANE}. 
To get initial embedding of network \bm{$Y^{(t)}_A$}, they solve a generalized eigen-problem $\mathbf{L}_{\mathrm{A}}^{(t)} \mathrm{a}=\lambda \mathbf{D}_{\mathrm{A}}^{(t)} \mathrm{a}$, where $a$ is the eigenvector and $\mathbf{Y}_{\mathrm{A}}^{(t)}=\left[\mathrm{a}_{2}, \ldots, \mathrm{a}_{k}, \mathrm{a}_{k+1}\right]$.
The initial embedding of  attributes \bm{$Y^{(t)}_X$} is obtained in the same way.
In order to resolve noisy data problem, they maximize the two embedding's correlation to get a consensus embedding \bm{$Y^t$}.
Furthermore, to capture dynamic information, at snapshot $t$, the variation of the eigenvalue $\lambda_{i}$ is derived as follows:
\begin{equation}
    \Delta \lambda_{i}=\mathrm{a}_{i}^{\prime} \Delta \mathrm{L}_{\mathrm{A}} \mathrm{a}_{i}-\lambda_{i} \mathrm{a}_{i}^{\prime} \Delta \mathrm{D}_{\mathrm{A}} \mathrm{a}_{i}.
\end{equation}
where \bm{$A^{t}$} is the adjacency matrix, $\Delta \mathrm{D}_{\mathrm{A}}$ and $\Delta \mathrm{L}_{\mathrm{A}}$ are the perturbation of the diagonal matrix and Laplacian matrix.

And the perturbation of eigenvector $a_i$ is given as follows:
\begin{equation}
    \begin{aligned}
    \Delta \mathrm{a}_{i}=&-\frac{1}{2} \mathrm{a}_{i}^{\prime} \Delta \mathrm{D}_{\mathrm{A}} \mathrm{a}_{i} \mathrm{a}_{i}\\
    &+\sum_{j=2, j \neq i}^{k+1}\left(\frac{\mathrm{a}_{j}^{\prime} \Delta \mathrm{L}_{\mathrm{A}} \mathrm{a}_{i}-\lambda_{i} \mathrm{a}_{j}^{\prime} \Delta \mathrm{D}_{\mathrm{A}} \mathrm{a}_{i}}{\lambda_{i}-\lambda_{j}}\right) \mathrm{a}_{j}
    \end{aligned}.
\end{equation}
the $i$-th eigen-pair $\left(\Delta \lambda_{i}, \Delta \mathbf{x}_{i}\right)$ of node attributes can also be updated in similar manner.

Similarly, \cite{DHPE} also adopts the generalized SVD to learn representation of network.
Specifically, they try to preserve high-order proximity by minimizing the following function:
\begin{equation}
    \min \left\|\mathbf{S}-\mathbf{U} \mathbf{U}^{\prime \top}\right\|_{F}^{2}.
\end{equation}
, where $\mathbf{U}, \mathbf{U}^{\prime} \in \mathcal{R}^{N \times d}$. In matrix decomposition, $\mathbf{U}$ and $\mathbf{U^{\prime}}$ can be seen as the basis and the coordinate. $S$ is Katz Index, which is used to measure high order proximity.
For incremental updating embedding results, they propose a generalized eigen perturbation to calculate the change of the eigenvalues and eigenvectors.

Though incremental matrix factorization methods could update the embedding results, \cite{Timers} observes that such approximations\cite{DANE} would lead to accumulated errors.
Therefore, they design a approach to optimally set the restart time of model by deriving a lower bound of SVD minimum loss and setting a maximum tolerated error as threshold.

Moreover, \cite{dyhne} proposes \textit{DyHNE} to learn node embeddings on heterogenous network. Specifically, they introduce meta-path based first- and second-order proximities to obtain heterogenous structural information.
And following \cite{DHPE}, \textit{DyHNE} also updates learned node embeddings by employing eigenvalue perturbation.

In summary, matrix factorization based methods model dynamic network in discrete way and learn node embedding by solving generalized eigen-problem.
Thanks to the matrix factorization mechanism, this type of model can flexibly handle the increase and decrease of nodes/edges.
Besides, this type of model is usually based on matrix perturbation to capture dynamics, which means it is suitable for dynamic graphs with subtle changes.
In addition, to reduce the computing complexity during matrix perturbation process, accelerated solutions need to be developed.

\subsubsection{Autoencoder based model}
\begin{figure}[ht]
    \centering
    \includegraphics[width = .45\textwidth]{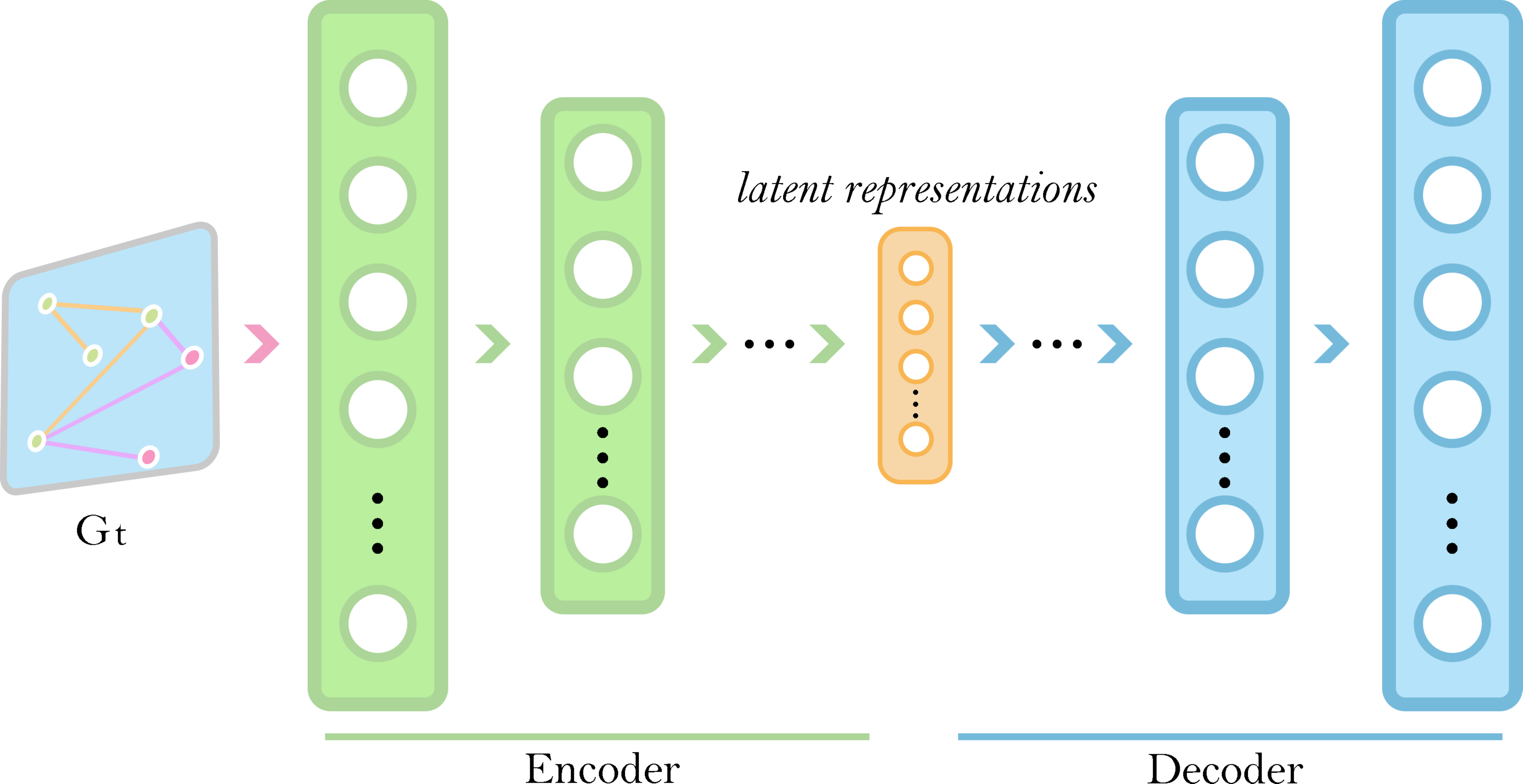}
    \caption{An illustration of autoencoder mechanism.}
    \label{pic:ae}
\end{figure}

The idea of autoencoder was firstly introduced in \cite{bourlard1988auto}. 
Figure~\ref{pic:ae} provides a simply illustration of autoencoder mechanism.
The encoder maps input samples to the feature space and generates latent representations while the decoder map them back to obtain the reconstructed samples.

We could learn embeddings by minimizing the reconstruction error. Most of autoencoders are implemented through various types of neural network.

Representing the dynamic graphs as a collection of snapshots, \textit{DynGEM} \cite{DynGEM} employs a deep autoencoder as its core and incrementally learn the embedding of snapshot $t$ from the embedding of previous snapshot $t-1$.
Specifically, \textit{DynGEM} initializes embedding at time $t$ with the learned embedding from time $t-1$, then carries out model training.
To work with growing number of nodes, they bring up a heuristic method to calculate the new sizes of neural network layers at each time step and add new layers if needed.
The loss function is aimed to preserve structural proximity with following form:
\begin
    {equation}L_{n e t}=L_{g l o b}+\alpha L_{l o c}+\nu_{1} L_{1}+\nu_{2} L_{2}.
\end{equation}
$L_{l o c} = \sum_{i, j}^{n} a_{i j}\left\|\boldsymbol{y}_{\boldsymbol{i}}-\boldsymbol{y}_{\boldsymbol{j}}\right\|_{2}^{2}$ is the first-order proximity, where $y$ is embedding value and $A$ is adjacency matrix.
And $L_{g l o b}=\sum_{i=1}^{n} \|\left(\hat{\boldsymbol{x}}_{\boldsymbol{i}}-\right.\left.\boldsymbol{x}_{\boldsymbol{i}}\right) \odot \boldsymbol{b}_{\boldsymbol{i}}\left\|_{2}^{2}=\right\|(\hat{X}-X) \odot B \|_{F}^{2}$
is the second-order proximity which is preserved by an unsupervised reconstruction of the neighborhood of each node.
$b_i$ is a vector with $b_{ij} = 1 $ if $a_{ij} = 0$ else $b_{ij} = \beta > 1$.

Instead of using network as input, \textit{NetWalk} \cite{Netwalk} takes a lists of vertices sampled by random walk.
Considering the sparsity of the input and output vectors, which are one-hot encoded, \textit{Netwalk} applies a sparse autoencoder to learn the embedding.
To be specific, KL-divergence is included as sparsity constraint in cost function, which measures the difference between two probability distributions.
In order to learn the network structure, they also design a loss term to minimize the pairwise distance among all vertices of each random walk.

\begin{figure}[ht]
    \centering
    \includegraphics[width = .5\textwidth]{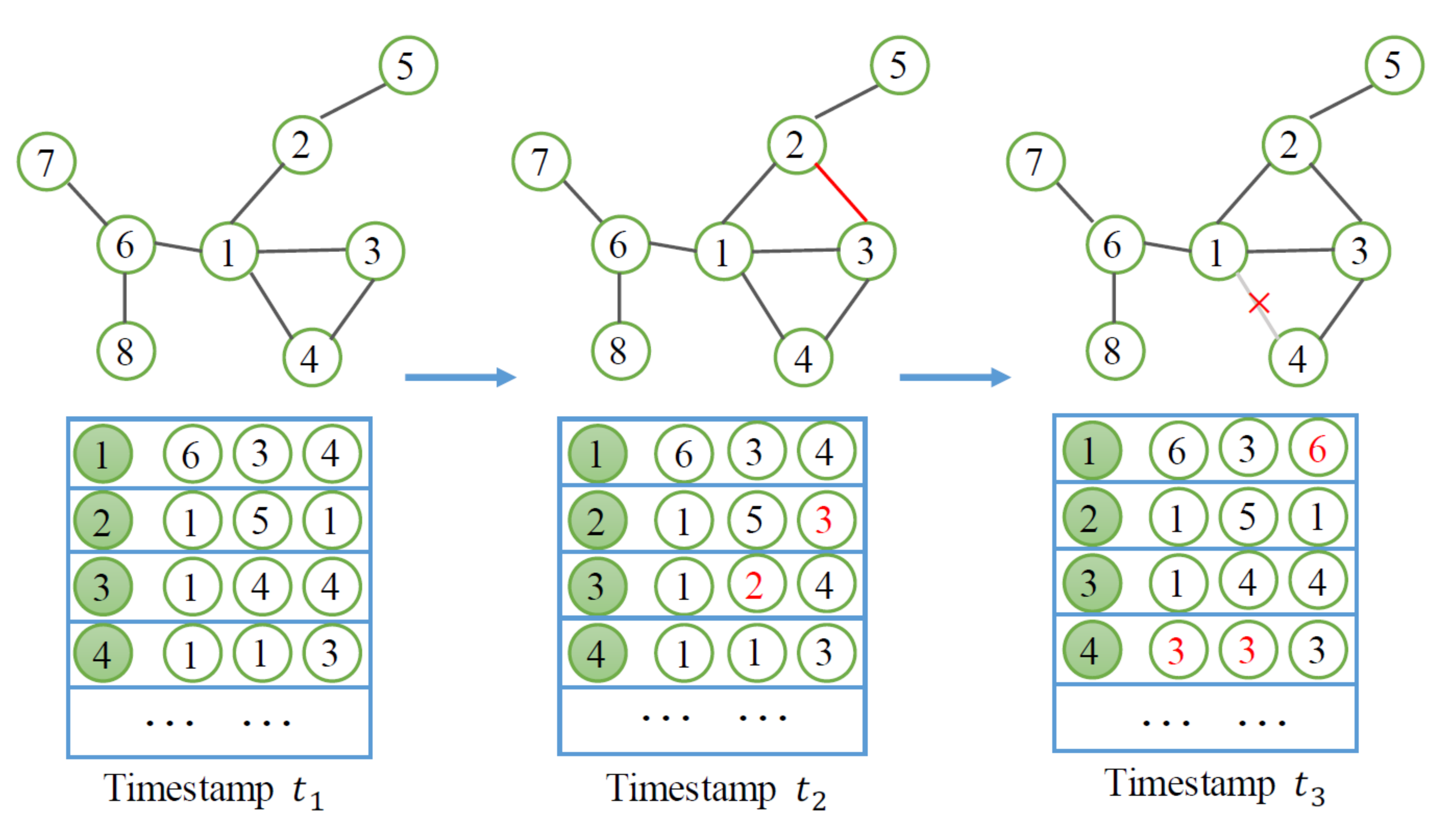}
    \caption{An illustration of updating the reservoirs. Figure extracted from \cite{Netwalk}}
    \label{pic:netwalk}
\end{figure}

\textit{Netwalk} treats the edges addition/deletion in a continuous view, which means the changes come like a stream.
Thus, \textit{Netwalk} maintains a reservoir to save random walk results for each vertex, which is showed in figure~\ref{pic:netwalk}.
The content of reservoir would update with probability when edge changing occurs.
Then they continue to train the model with the updated network walk set in a warm-start fashion.

\cite{dyngraph2vec} observes that \textit{DynGEM}, \cite{DynamicTriad} and \textit{Timers} only utilize the previous time step network, which imply they assume the patterns are of short duration.
Besides, \textit{DynGEM} and \textit{Timers} also assume that the changes are smooth and impose a regularizer to ensure temporal smoothness.
Based on the above observations, \textit{dyngraph2vec} hopes to use longer-term historical information to learn node embeddings. Therefore, \textit{dyngraph2vec} takes $l$ snapshots as input and predict the $t+l+1$ network. It integrates dense layer and LSTM layer as encoder to capture dynamic information.
Moreover, \cite{BurstGraph} based on observation of user buying behavior, propose \textit{BurstGraph} to especially capture bursty network changes.
\textit{BurstGraph} uses \textit{GraphSage} \cite{graphsage} as encoder while utilizes RNN as a part of decoder to tackle with the evolution of network.
Specifically, \textit{BurstGraph} introduces a spike-and-slab distribution as an approximation of posterior distribution in the variational autoencoder framework\cite{VAE} to deal with bursty links.

In short, the above methods show that the autoencoder as a general framework can learn the node representation on the dynamic network.
The main difference between the methods is how to better capture temporal information. In particular, it is necessary to design different encoder according to different network change modes, such as smooth changes.

\subsubsection{Skip-gram based model}
Introduced in \textit{Word2Vec}\cite{goldberg2014word2vec}, Skip-gram is a simple but effective model to learn embedding representations.
It's originally applied in word embedding task, which uses a input word to predict corresponding context.
\textit{DeepWalk} \cite{perozzi2014deepwalk} generates sequences of vertices by random walk. Then, by treating walks as the equivalent of sentences, \textit{DeepWalk} utilized Skip-gram model to learn node embeddings.
\textit{Node2Vec}\cite{grover2016node2vec}, \textit{LINE}\cite{tang2015line} and other models make further progress to better capture network structure.

To generalize Skip-gram based network embedding techniques to dynamic field, \cite{extendedskipgram} propose a decomposable objective equivalent to the objective of \textit{LINE}.
The adjusted objective function could learn latent representations for new vertices in snapshot.
Besides, \cite{extendedskipgram} also analyzes the influence of network evolution and design a selection mechanism to choose the greatly affected nodes to update.

Also tackling with a sequence of snapshots, \cite{dynnode2vec} modified \textit{Node2Vec} to learn dynamic network embedding.
As repeatedly generating random walk for every snapshot is time-consuming, \textit{dynnode2vec} only update the list of random walk for evolving nodes.
The evolving nodes in the timestamp $t$ are defined as:
$$\Delta V_{t}=V_{a d d} \cup\left\{v_{i} \in V_{t} \mid \exists e_{i}=\left(v_{i}, v_{j}\right) \in\left(E_{a d d} \cup E_{d e l}\right)\right\}.$$
Similarly to \textit{DynGEM}\cite{DynGEM}, \textit{dynnode2vec} inherits parameters' weight from previous model(snapshot $t-1$), and train the current model with updated random walks.

\begin{figure}
    \centering
    \includegraphics[width = .4\textwidth]{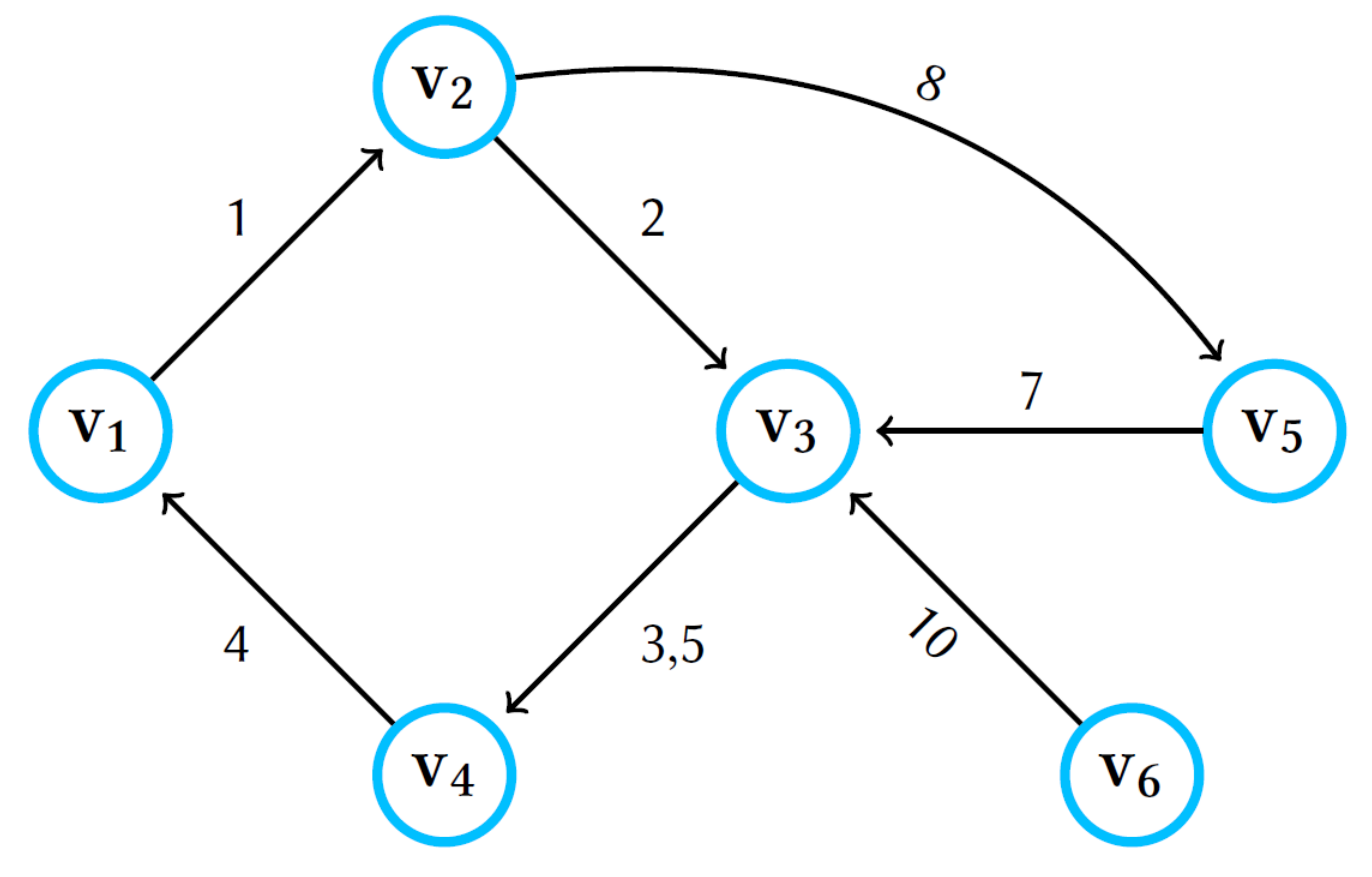}
    \caption{$v_2$,$v_3$,$v_4$ is a valid walk while $v_5$,$v_3$,$v_4$ isn't a valid walk since edge between $v_5$,$v_3$ is created later than edge between $v_4$,$v_3$. Figure extracted from \cite{ContinuoustimeDNE}}
    \label{pic:continuousDNE}
\end{figure}

However, \cite{ContinuoustimeDNE} argues that representing dynamic network as a series of snapshots leads to coarse and noisy approximation. They suggest to model the temporal network as a continuous model.
Such modeling approach may not only reduce the loss of information, but also let researchers could flexibly select appropriate granularity. That is to say researchers could customize the time interval.
To be specific, as all the coming edges are annotated with timestamps, \cite{ContinuoustimeDNE} add temporal constraint to random walk. The walk should traverse along edges in the direction of increasing time, which is illustrated at figure~\ref{pic:continuousDNE}.
In order to perform temporal random walks, the selection of initial edges and time has also been discussed. In addition, with the assumption that the nodes with new arriving edges are impacted by historical events, \textit{HTNE}\cite{htne} applies Hawkes process to model the neighborhood formation.

In summary, Skip-gram based methods are often based on random walks to design sampling methods, thereby updating the context information of the node in network evolution.
Benefiting from updating only a subset of nodes, this type of method exhibits higher performance when the network structure changes little.

\subsubsection{GNN based model}
Recently, applying neural network on network has made a booming development. Several kinds of graph neural network have been proposed and achieved impressive results on relevant tasks.
Most of the models extend deep learning approaches for network data. For example, \textit{GCNs} \cite{bruna2013spectral,kipf2016semi,niepert2016learning, defferrard2016convolutional,graphsage} realize convolution-like operation on graphs,
attention mechanism is also applied in graph neural network\cite{velivckovic2017graph,wang2019heterogeneous}.

\textit{STGCN}\cite{stgcn} focuses on solving the traffic prediction problem on traffic graphs.
Due to all the positions of monitor station are fixed, a dynamic traffic network has a constant size of nodes.
As time goes by, the node's features would keep changing, which reflect the traffic condition.
\textit{STGCN} adopts the first-order approximation of spectral graph convolutions\cite{kipf2016semi} to gather the spatial information.
Hoping to respond quickly to dynamic changes, instead of using RNN-based architecture, \textit{STGCN} design a temporal convolution layer to capture dynamic behaviors of traffic flows.
It integrates 1-D causal convolution and gated linear units.

However, network model in \textit{STGCN} is a special case of dynamic network without growing size. \cite{dysat} proposes \textit{DySAT}, which is able to learn embedding from more general form of dynamic network.
Observed that previous approaches\cite{DynamicTriad,DANE}, which pursue smoothness of node representations, may ignore distinct temporal behaviors, \cite{dysat} introduces self-attention mechanism.
It has a clear architecture: model consists of a structural attention block and a temporal attention block.
In structural block, \textit{GAT}\cite{velivckovic2017graph} is modified to collect spatial information for each snapshot and output a sequence of node embeddings at different time steps.
Every node's embedding sequence is push forward to temporal attention block to learn the final representation.
Benefiting by flexible attention weight, model could capture temporal dependencies at a fine-grained node-level granularity, rather than output smooth expressions for nodes.
Besides, inspired by \textit{DySAT}, \textit{DyHAN}\cite{dyhan} uses hierarchical attention to learn node embedding, which performs better than \textit{DySAT} experiment.
It is worth mentioning that \textit{DyHAN} also tackles with the structural heterogeneity of network. And experimental results show the efficacy of considering heterogeneity.

Since graph neural network could only gather neighborhood's features and provide embedding for static network, RNN-like techniques are also integrated to learn temporal information in some models.
However, it's doubtful that whether RNNs could learn a valid embedding for new coming nodes in consideration of lacking historical features.
In order to resolve this challenge while make use of RNNs to capture dynamic information at the same time, \cite{Evolvegcn} proposes \textit{EvolveGCN}.
Basically, it uses GCN to generate node embeddings for every snapshot. Then, instead of directly applying recurrent layer to refine the embeddings, \textit{EvolveGCN} uses RNN to train the weights of GCN.
It provides GRU version and LSTM version of model, which could be written as:
\begin{equation}
    \overbrace{\underbrace{W^{(l)}_t}_{hidden\ state}}^{GCN\ weights} = GRU(\overbrace{\underbrace{H^{(l)}_t}_{input}}^{node\ embeddings},\overbrace{\underbrace{W^{(l)}_{t-1}}_{hidden\ state}}^{GCN\ weights}),
\end{equation}

\begin{equation}
    \overbrace{\underbrace{W^{(l)}_t}_{output}}^{GCN\ weights} = LSTM(\overbrace{\underbrace{W^{(l)}_{t-1}}_{input}}^{GCN\ weights}).
\end{equation}
Though it is fancy to maintain temporal information through evolving parameters, \textit{EvolveGCN}'s performance in link prediction is not outstanding.

Focusing on anomaly detection, \cite{AddGraph} considers that \textit{NetWalk}\cite{Netwalk} fails to capture the long-term and short-term patterns.
Therefore, \cite{AddGraph} propose \textit{AddGraph}, which also integrates RNN as part of model.
At time $t$, current state representation is propagated by GCN,
\begin{equation}
    \text { Current }^{t}=\operatorname{GCN}_{L}\left(\mathbf{H}^{t-1}\right).
\end{equation}
, where $\mathbf{H^{t-1}}$ is hidden state learned at time $t-1$.
To collect more historical information, a contextual attention-based model\cite{cui2019hierarchical} is applied.
The model could be briefly written as a function:
\begin{equation}
    \operatorname{Short}^{t}=\operatorname{CAB}\left(\mathbf{H}^{t-\omega} ; \ldots ; \mathbf{H}^{t-1}\right)
\end{equation}
, where $\omega$ is the size of window.
Then, \textit{AddGraph} applies GRU to encode current state and historical resources as:
\begin{equation}
    \mathbf{H}^{t}=\operatorname{GRU}\left(\text { Current }^{t}, \text { Short }^{t}\right)
\end{equation}.
As labeled data is especially insufficient in anomaly detection, negative sampling strategy and margin loss technique are also applied.
What's more, other dynamic network embedding models\cite{DGNN, BurstGraph} also integrate GNN approaches to better extract structural features.

In general, unlike previous methods, the methods based on graph neural network can easily process attributed dynamic network.
The difference is the strategy of capturing the network evolution. Some methods use parameter reuse to implicitly preserve temporal information, while others use attention mechanisms or RNNs to explicitly encode temporal information.
In addition, because such methods are updated based on snapshots, it is not easy to perform more fine-grained and more timely node representation learning.

\subsection{Temporal-first model}
\subsubsection{RNN based model}
Based on \cite{rumelhart1986learning}, recurrent neural network(RNN) is very general form of neural network.
RNN is designed for processing sequential data, where connections between neurons form a directed network along a temporal sequence.
Outstanding models like GRU\cite{chung2014empirical}, LSTM\cite{hochreiter1997long}, use neuron's internal state as memory to process variable length sequences of inputs.

As RNN's intrinsic ability to encode temporal information, simple RNN units are adopted to learn entity embedding over dynamic knowledge graph\cite{knowevolve}.
\textit{DGNN}\cite{DGNN} try to learn the global impact of new edges while also consider local influence for new connected nodes.
In essence, \textit{DGNN} wants to update embedding of source node and endpoint when new interaction occur. Besides, the influence should also propagate to the first-order neighbors.
Based on above motivations, it makes extension of LSTM, which can take embeddings of first-order neighbor and endpoint of a edge as input respectively.

Similarly, based on the observation of social network, \cite{Dyrep} points out that the evolution of network contains two fundamental processes: association process and communication process.
The former one represents the topology change which establishes or destroys a long-lasting bridge for information exchange between nodes, while the latter process means information propagates for a short time at adjacent nodes.
To model the processes above, \textit{DyRep} model dynamic network in continuous way and implement a RNN-like structure with attention mechanism.
Instead of directly give embedding as output, the model learns functions to compute node representations, which is capable of inferring representation for unseen node.

\cite{jodie} also adopts the continuous model. In particular, the model \textit{JODIE} is customized for User-Item bipartite network, which is a common pattern in recommendation system.
\textit{JODIE} builds up a coupled RNN architecture and adds a attention layer to get final representation of user.
It uses two separate RNN to update the embeddings of user and item for every interaction occurs.
Take the update of user embedding as example, it could be formally written as:
\begin{equation}
    u(t)=\sigma\left(W_{1}^{u} u\left(t^{-}\right)+W_{2}^{u} i\left(t^{-}\right)+W_{3}^{u} f+W_{4}^{u} \Delta_{u}\right).
\end{equation}
, where $u(t^-)$, $i(t^-)$ are previous embedding of user and item, $f$ is embedding of static features and $\Delta_u$ denotes the time since u’s previous interaction.
In this way, \textit{JODIE} makes use of  stationary properties and dynamic information in a unified framework.
What's more, according to life experience, user's intent would not stay constant over time even no interaction with item occurred.
This requires the updating embeddings could predict the trajectory of moving intent when there is no fresh information of nodes.
In fact, the prediction of trajectory could be considered as embedding projection. And a linear operator is proposed to accomplish it.
Therefore, the final embedding of user at time $t+\Delta$ is predicted as:
\begin{equation}
    \begin{aligned}
        \tilde{j}(t+\Delta)=&W_{1} \widehat{u}(t+\Delta)+W_{2} \bar{u}\\
        &+W_{3} i\left(t+\Delta^{-}\right)+W_{4} \bar{i}+B.
    \end{aligned}
\end{equation}
, where $\widehat{u}(t+\Delta)$, $i\left(t+\Delta^{-}\right)$ are the projected user embedding and item embedding.

In addition, heterogeneous dynamic graphs have also received more and more attention. 
\cite{DyHATR} proposes \textit{DyHATR}, which uses the hierarchical attention model to capture the heterogeneity and introduces the temporal attentive GRU/LSTM to model the evolutionary patterns among snapshots.
While \cite{THIGE} focus on next-item recommendation problem, they learn embedings on temporal heterogenous User-Item bipartite network. This kind of network has multiple types of edges, representing different interaction behaviors.
Specifically, they divide the user's historical interactions into long-term interactions and short-term interactions based on timestamps. They introduce heterogeneous self-attention mechanism to learn long-term interaction while use GRU to learn users' current demands.

In short, RNN based methods usually adopt continuous model, which can update node and edge changes more timely. This type of method is usually used in recommendation systems.
The main difference between each method is how to design a node representation learning method according to a specific network structure, such as a User-Item bipartite network.

\subsection{Others}
There are some works learn the dynamic network embedding in their unique ways.
Facing the challenge of inferring embedding for out-of-sample nodes, \textit{DepthLGP} is proposed by \cite{DepthLGP}.
They design a high-order Laplacian Gaussian process (hLGP) to encode network properties and use a deep neural network to perform nonlinear transformation.
Benefiting from nonparametric probabilistic modeling, \textit{DepthLGP} maintains high speed for scalable inference.

Besides, \textit{DynamicTriad}\cite{DynamicTriad}, a semi-supervised algorithm, learns embeddings by modeling the triadic closure process.
Triadic closure is the property among three nodes, where an open triad is two of three nodes have connections and a closed triad means connection exists between any two nodes.
They define the probability that the open triad $(vi, vj, vk)$ evolves into a closed triad as
\begin{equation}
    P_{\mathrm{tr}}^{t}(i, j, k)=\frac{1}{1+\exp \left(-\left\langle\boldsymbol{\theta}, \boldsymbol{x}_{i j k}^{t}\right\rangle\right)},
\end{equation}
where $x^t_{ijk}$ is vector calculated using embeddings of node $i,j,k$, and $\theta$ is a social strategy parameter.
In order to learn node embeddings, corresponding loss function and sampling strategy are proposed.

Moreover, using continuous data model, \cite{MMDNE} studies network dynamics from the micro and macro perspectives and proposes \textit{MMDNE}. 
Micro-dynamics describes the specific process of network structure changes, such as newly established edges, while macro-dynamics describes the evolution pattern of network scale.
Specifically, they introduce temporal point process to describe the establishment of edges, i.e. micro-dynamics, and combine the hierarchical temporal attention to learn node representations.
And for macro-dynamics, they define a general dynamics equation parameterized with network embeddings

In addition, in order to generate representations in an inductive fashion, temporal graph attention(\textit{TGAT})\cite{tgat} proposes a functional time encoding technique by applying Bochner’s theorem.
Self-attention mechanism is also been applied as building block to aggregate neighborhood information.

\section{Discussion}
\label{sec:discussion}
In this section, we discuss several critical challenges in learning dynamic network embeddings and correspondingly proposed methods.

\begin{itemize}
    \item
    \textbf{Long-term Features Preservation}

    In the process of network evolution, special patterns such as communities and interactive features may be formed, which are worth mining and preserving.
    For example, a forming hub may indicate illegal transactions in financial network.
    And in user-item network, a user frequently interacts with an item could serve as a long-term feature which reflect his/her love for certain classes of products.
    Some models are explicitly modeled to preserve long-term features, while others implicitly learn these features when node embeddings are updated.
    Based on the assumption that network changes smoothly, most matrix factorization based methods adopt matrix perturbation\cite{stewart1990matrix} theory to update the embedding results\cite{DHPE,DANE,Timers}.
    For deep learning models, reusing parameters from previous time for initialization\cite{DynGEM,dynnode2vec} is one of the directly way to preserve dynamic information.
    However, weights reusing does not explicitly pay attention to long-term features Preservation.
    By maintaining sequences of nodes at different time steps, many methods\cite{DGNN,jodie,Evolvegcn,AddGraph,dyngraph2vec} introduce RNN module, especially, GRU and LSTM, to particularly preserve long-term features while absorb current features in the same time.
    Besides, Hawkes process\cite{htne} and attention mechanism\cite{dysat,tgat,dyhan} are also introduced to model dynamic features.

    \item
    \textbf{Out-of-sample Node Embedding}

    As many classical static embedding methods are transductive algorithm\cite{perozzi2014deepwalk,kipf2016semi}, they could not infer embeddings for out-of-sample nodes.
    However, most of dynamic network embedding methods, which take node addition/deletion into account, can continue model training and learn embeddings for out-of-sample nodes.
    Because out-of-sample nodes can be naturally treated as a part of network in next time step.
    Many classical static embedding methods are transductive algorithm\cite{perozzi2014deepwalk,kipf2016semi}, which could not infer embeddings for out-of-sample nodes.
    However, learning node embedding on dynamic network provide a more natural way to think of out-of-sample nodes. Because those nodes can be naturally treated as a part of network in next time step. 
    Besides, most of dynamic network embedding methods, which take node addition/deletion into account, can continue model training and learn embeddings for out-of-sample nodes. 
    Further more, inductive learning methods on dynamic network leverage temporal features, which could bring more useful information than inductive learning methods on static network\cite{graphsage,velivckovic2017graph, rossi2018deep, bai2019unsupervised}.

    Also, in order to improve the efficiency of incrementally model training for out-of-sample nodes, lots of tricks have been used.
    For example, \textit{dynnode2vec}\cite{dynnode2vec} and \textit{NetWalk}\cite{Netwalk} only generate random walks for changed nodes while \cite{jodie} proposes \textit{t-Batch} algorithm to accelerate training.
    Furthermore, \textit{DyRep}\cite{Dyrep} and \textit{TGAT}\cite{tgat} propose inductive algorithm that could learn representations for unseen nodes directly.

    \item
    \textbf{Prediction for Future Embedding}

    Predicting future representations is really helpful for practical applications, such as products recommendation, fraud detection.
    A few of models try to predict the representation of the next time step.
    \textit{dyngraph2vec}\cite{dyngraph2vec} utilizes autoencoder structure by reducing reconstruction loss, which generates node embedding prediction for next snapshot.
    Besides, \cite{jodie} considers that user's interest may change when they don't interact with items.
    Motivated by this, \cite{jodie} designs a linear operator to predict the future trajectory of user in the embedding space.
    While \textit{STGCN}\cite{stgcn} integrates gated CNN and GCN to give traffic prediction.

\end{itemize}

\section{Data Sets, Evaluation and Applications}
\label{sec:application}
In this section, we summarize the datasets and evaluation tasks that are commonly used in developing dynamic network embedding methods.
Besides, we also describe practical applications in various domains.

\subsection{Datasets}
We summarize commonly used dynamic network dataset with its size and granularity in table~\ref{table:dataset}.

\begin{table*}[!t]
\begin{threeparttable}
    \centering
    \caption{Commonly used dataset}
    \footnotesize
    \label{table:dataset}
    \begin{tabular}{|m{2.5cm}<{\centering}|m{9cm}|m{1.5cm}<{\centering}|m{2cm}<{\centering}|m{1.8cm}<{\centering}|}
        \hline
        Name&Description&Nodes&Edges&Granularity\\
        \hline
        AS-733\cite{leskovec2005graphs}&This dataset contains traffic flow exchanging in Autonomous Systems. It exhibits both the addition and deletion of the nodes and edges over time.
        &6,474&13,895&Daily\\
        \hline
        CollegeMsg\cite{panzarasa2009patterns}&This dataset is a directed network comprised of private messages sent on an online social network at the University of California, Irvine.
        &1,899&\makecell[c]{T: 59,835\\S: 20,296}&Seconds\\
        \hline
        DBLP\cite{yang2015defining}&This is a co-authorship network where two authors are connected if they publish at least one paper together.
        &317,080&1,049,866&-\\
        \hline
        wiki-talk\cite{paranjape2017motifs}&This dataset is a network representing Wikipedia users editing each other's Talk page. A directed edge $(u, v, t)$ means that user $u$ edited user $v$'s talk page at time $t$.
        &1,140,149&\makecell[c]{T: 7,833,140\\S: 3,309,592}&Seconds\\
        \hline
        Enron\cite{klimt2004introducing}&This dataset contains email messages from the Enron corpus and constructs a E-mail communication network.
        &143&2,347&Seconds\\
        \hline
        HEP-TH\cite{leskovec2005graphs}&This is a citation network from the e-print arXiv., which covers papers in the period from January 1993 to April 2003 (124 months).
        &27770&352807&Monthly\\
        \hline
        Epinions\cite{richardson2003trust}&This is a who-trust-whom online social network of a a general consumer review site Epinions.com.
        &75,879&508,837&-\\
        \hline
        Reddit\cite{kumar2018community}&This is a subreddit-to-subreddit hyperlink network which is extracted from the posts that create hyperlinks from one subreddit to another.
        &55,863&858,490&Seconds\\
        \hline
        Elliptic\cite{weber2019anti}&This is a transaction network collected from the Bitcoin blockchain. A node represents a transaction, while an edge can be viewed as a flow of Bitcoins between one transaction and the other.
        &203,769&234,355&49 time steps\\
        \hline
    \end{tabular}

    \begin{tablenotes}
        \item \textbf{T} means the number of temporal edges while \textbf{S} means the number of static edges.
    \end{tablenotes}
\end{threeparttable}
\end{table*}

\subsection{Evaluations}
We briefly introduce two often used tasks in network embedding. We also list other tasks used in dynamic network embedding in table~\ref{table:summary}.

\subsubsection{Node Classification}
In most of graphs, a portion of nodes are assigned with labels. For example, in citation networks, a node may be labeled with its research field, whereas the labels of nodes in E-commerce may based on user's interest.
Due to various factors, labels may be unknown for large fractions of nodes.  The goal of node classification is to assign each node with appropriate label by learned features.
Generally speaking, the process of node classification based on the network embedding could be summarized as follow:
(1) Obtain low dimensional features with the network embedding approaches.
(2) Train a classifier with learned embeddings to divide nodes into corresponding categories.
Generally, macro-F1 and AUC (Area Under Curve) are used to evaluate the performance.

\subsubsection{Link Prediction}
Link prediction is also a fundamental problem in network analysis. Graphs are often incomplete in real world, e.g., links could be missing for two friends in social network.
Also, researchers are interested in predicting network topology in the future. Hence, link prediction refers to predict either missing links or edges that may generate in the future.
Essentially, it aims to predict whether there exists an edge between two nodes, which could be considered as a binary classification problem.
Therefore, labels indicate whether there is a connection for a pair of nodes. Generally, precision @k and Mean Average Precision (MAP) are often used metrics


\subsection{Practical applications}
In addition to fundamental tasks listed above, dynamic network embedding methods also been adopted to practical applications across different domains.
We briefly describe serval applications which handle network structure data with dynamic embedding techniques.

\begin{itemize}
    \item
    \textbf{Anomaly Detection}

    Anomaly detection aims to identify the anomalous nodes in network, whose behaviors are different from the common nodes.
    And dynamic network is more likely to expose defaulter's malicious pattern.
    For example, in financial network, a deceiver may pretend to be a normal user in every deal but its transaction pattern through timeline would be abnormal.
    Early fraud detection is crucial to the minimization of loss to a financial institute and the protection of normal user rights.
    Anomaly detection is also important for defending network attack, detecting ticket scalper in E-commerce, etc..
    Serval works have utilized dynamic network to perform detection. \cite{AddGraph,Netwalk,yang2020h,garchery2020adsage,lagraa2020simple}
    \item
    \textbf{Recommender System}

    Network-based recommender systems usually take items and users as nodes and build up connections between items and items, users and users, items and users.
    Besides, temporal network provides fine-grained granularity of history which contains more information than static network.
    The dynamic network embedding methods serve as a powerful way to reduce dimension and extract items or users features, which facilitate the downstream applications.
    Furthermore, as the pivotal issue of recommender system is scoring the importance of an item to user, we can consider it as a link prediction problem.
    That is, whether we recommend a item to a user depends on whether a missing links is predicted by applied model.
    \cite{BurstGraph} proposes \textit{BurstGraph} to capture unexpected bursty changes and \cite{jodie} uses a coupled RNN to update embeddings for users and items.
    \cite{DGRec} completes social recommendation based on session while \cite{THIGE} completes next-item recommendation on temporal heterogeneous network. 
    \item
    \textbf{Traffic Forecasting}

    As traffic flow can naturally be modeled as a directed network, several models of dynamic network embedding also attend to predict traffic.
    \cite{stgcn} firstly adopts graph convolutional network, \cite{li2017diffusion} models traffic flows as a diffusion process and \cite{geng2019spatiotemporal} focus on ride hailing demand forecasting.
    In addition, \cite{lu2020st,song2020spatial,peng2020spatial} are also studying traffic forecasting task.
\end{itemize}

\begin{table*}[!t]
    \centering
    \caption{A Summary of the Source Code}
    \label{table:codes}
    \begin{tabular}{@{}cc@{}}
    \toprule
    \multicolumn{2}{c}{\textbf{Structural Model}} \\ \midrule
    Model & Source code \\ \midrule
    DyHNE & https://github.com/rootlu/DyHNE \\
    TIMER &  http://nrl.thumedia.org/timers-error-bounded-svd-restart-on-dynamic-networks \\
    CTDN & https://github.com/LogicJake/CTDNE \\
    NetWalk & https://github.com/chengw07/NetWalk \\
    BurstGraph & https://github.com/ericZhao93/BurstGraph \\
    DySAT & https://github.com/aravindsankar28/DySAT \\
    STGCN & https://github.com/VeritasYin/STGCN\_IJCAI-18 \\
    EvolveGCN & https://github.com/IBM/EvolveGCN \\ \bottomrule
    \multicolumn{2}{c}{\textbf{Temporal Model}} \\ \midrule
    Model & Source code \\ \midrule 
    Know-Evolve & https://github.com/rstriv/Know-Evolve \\
    JODIE & http://snap.stanford.edu/jodie \\
    DyHATR & https://github.com/skx300/DyHATR \\
    THIGE & https://github.com/yuduo93/THIGE \\ \bottomrule
    \multicolumn{2}{c}{\textbf{Others}} \\ \midrule
    Model & Source code \\ \midrule
    DynamicTriad & https://github.com/luckiezhou/DynamicTriad \\
    TGAT & \makecell[c]{https://github.com/StatsDLMathsRecomSys/ \\ Inductive-representation-learning-on-temporal-graphs } \\ \bottomrule
    MMDNE & https://github.com/rootlu/MMDNE \\
    \multicolumn{2}{c}{\textbf{Toolkit}} \\ \midrule
    Name & Source code \\ \midrule
    DynamicGEM & https://github.com/palash1992/DynamicGEM \\ \bottomrule
    \end{tabular}
\end{table*}
    
\subsection{Open Source Software}
We collect a set of open source code links for dynamic network embedding methods in table~\ref{table:codes}.

\section{Conclusion and Future Direction}
In this survey, we conduct a comprehensive overview of the literatures in dynamic network embedding.
We first provide formal definitions of commonly used data model of dynamic network: discrete model and continuous model.
Then, according to the data models and corresponding methodologies, we propose a new taxonomy that organizes current dynamic network embedding methods within a novel category hierarchy, and present and compare them briefly.
Some critical challenges in dynamic network embedding are also discussed.
Lastly, we introduce useful dynamic network datasets and several real applications.

In the future, we believe there are at least four promising research directions for dynamic network embedding:
1) Learning embedding on heterogenous network. 
Heterogenous network has shown great potential to provide richer information with various node types and edge types.
The static embedding on heterogenous network is well studied recently~\cite{wang2019heterogeneous,zhang2019heterogeneous,fan2019metapath} while the research on heterogeneous dynamic graphs is just getting started\cite{dyhan,THIGE,DyHATR,bian2019network}.
How to further explore the relationship between various types of nodes and edges in the process of network change, and to mine the unique time patterns of heterogeneous graphs is a direction worth studying.
2) Considering more general structure changes. As we summarize in Table~\ref{table:summary}, many works do not consider the deletion of node or edge.
The changes of nodes and edges in the network imply a lot of information, such as may lead to the production of special subgraphs.
Supporting more general structure changes would better describe the evolution of network.
3) Applying model on large-scale network. The scalability of model should also be taken into account. 
As the network evolves, the scale of the network may gradually grow, leading to the accumulation of snapshots in discrete model or timestamps in continuous model.
How to reduce the complexity of the model to make it suitable for the ever-increasing network, while exploring more temporal information from the accumulated historical records, is a direction worthy of further study.
4) Designing task-oriented model. In addition to node classification and link prediction, there are many tasks in the field of network data mining to explore. 
As the network evolves, there may be dense connections between a subset of nodes, which means that a community structure is formed. 
How to perform representation learning on dynamic graphs for specialized tasks such as community detection is also worthy of in-depth study.
5) Exploring model interpretability. Lastly, focusing on designing interpretable model could be fruitful as current approaches are limited in interpretability.



\bibliographystyle{elsarticle-num}
\bibliography{ref}

\end{document}